\documentclass[conference]{IEEEtran}

\usepackage{cite}
\usepackage{graphicx}
\usepackage{url}
\usepackage{subcaption}
\usepackage{listings}
\usepackage{booktabs}
\usepackage{multicol}
\usepackage[multiple]{footmisc}
\usepackage{flushend}

\newcommand{\revealing}[1]{#1}

\newcommand{\qt}[1]{``#1''}
\newcommand{\code}[1]{\texttt{#1}}

\newcommand{\terminal}[1]{\texttt{#1}}
\newcommand{\nonterminal}[1]{\textit{#1}}
\newcommand{\optionalTerm}[1]{\textit{(}#1\textit{)?}}
\newcommand{\multiTerm}[1]{\textit{(}#1\textit{)*}}
\newcommand{\parenTerm}[1]{\textit{(}#1\textit{)}}
\newcommand{\ruleseparator}{\medskip\hrule\medskip}

\newcommand{\rulesep}{\unskip\ \vrule\ }

\lstdefinelanguage{apidef}{
	morekeywords={api,record,enum,optional,optin,replaces,as},
	sensitive=true,
	morecomment=[l]{//},
}

\begin{document}

\title{Continuous API Evolution in Heterogenous Enterprise Software Systems}

\author{\IEEEauthorblockN{\revealing{Holger Knoche}}
\IEEEauthorblockA{\textit{\revealing{Software Engineering Group}} \\
\textit{\revealing{Kiel University}}\\
\revealing{Kiel, Germany} \\
\revealing{hkn@informatik.uni-kiel.de}}
\and
\IEEEauthorblockN{\revealing{Wilhelm Hasselbring}}
\IEEEauthorblockA{\textit{\revealing{Software Engineering Group}} \\
\textit{\revealing{Kiel University}}\\
\revealing{Kiel, Germany} \\
\revealing{wha@informatik.uni-kiel.de}}
}

\maketitle

\begin{abstract}
The ability to independently deploy parts of a software system is one of the cornerstones of modern software development, and allows for these parts to evolve independently and at different speeds.

A major challenge of such independent deployment, however, is to ensure that despite their individual evolution, the interfaces between interacting parts remain compatible.
This is especially important for enterprise software systems, which are often highly integrated and based on heterogenous IT infrastructures.

Although several approaches for interface evolution have been proposed, many of these rely on the developer to adhere to certain rules, but provide little guidance for doing so.
In this paper, we present an approach for interface evolution that is easy to use for developers, and also addresses typical challenges of heterogenous enterprise software, especially legacy system integration.

\end{abstract}

\section{Introduction}
\noindent Many modern approaches to software engineering and delivery rely on the ability to deploy parts of a software system independently of each other.
For instance, Newman \cite{Newman2020} highlights independent deployability as one of the key advantages of microservices.
Practices like Continous Delivery \cite{HumbleFarley2011}, which aim at fast feedback from production to development, would be impossible to implement without this ability.

In recent years, we have observed that also more conservative companies, such as banks and insurance companies, have begun to embrace such practices for their internal software systems \cite{EMISA2019}, in particular for migrating legacy systems towards microservices \cite{IEEESoftware2018}.
There are several reasons for this development; in our case study from the insurance domain, which we will present later on in further detail, the key drivers are the desire to shorten the time-to-market, implement agile practices, and allow applications to evolve at different speeds.

However, said applications are often highly integrated, and interact with each other in complex ways.
One of the key challenges is therefore to ensure that the applications remain compatible with each other at all times, even when they are independently deployed.
Furthermore, some of these applications are several decades old and based on aged technologies or platforms.
For such applications, libraries and utilities that are commonplace on modern platforms may not be available.
Therefore, approaches are needed that can be implemented even on dated implementation platforms with acceptable effort.

In this paper, we present an approach that allows to continuously evolve the interfaces of a software application while ensuring the compatiblity with dependent applications.
In particular, we aim to address the following goals:

\begin{enumerate}
	\item The approach should be easy to use for the developers, and help to avoid common mistakes
	
	\item The approach should have as little impact as possible on the developer's day-to-day work
	
	\item The approach must also support aged implementation technologies and platforms
	
	\item The runtime performance impact of the approach must be acceptable
\end{enumerate}

\noindent The remainder of this paper is structured as follows.
In Section~\ref{sec:case-study}, we introduce the case study that motivated this work.
An evolution example is given in Section~\ref{sec:example}.
Existing approaches to API evolution are discussed in Section~\ref{sec:existing-approaches}.
Our approach, which constitutes the main contribution of this paper, is presented in Section~\ref{sec:approach}, and a short evaluation is described in Section~\ref{sec:implementation}.
Related work is discussed in Section~\ref{sec:related-work}, and Section~\ref{sec:conclusions} concludes the paper.
\section{Case Study}
\label{sec:case-study}

\noindent The underlying case study for this work is the core insurance software system of a German insurance company.
The software system has in large parts been developed by the company itself, and consists of more than 30 individual, but highly integrated applications.
Development of the first applications started in the 1970s in COBOL on mainframe computers, and even today, the vast majority of the codebase (about 14 million lines of code) is COBOL code.
In the early 2000s, Java EE was adopted as a new implementation platform, and newer applications have been developed solely in Java.
As of today, these applications amount to about 2 million lines of code.

In order to establish cross-platform standards, the company adopted Model-Driven Software Development (MDSD) \cite{StahlVoelter2006} in the 2000s.
A considerable amount of code is generated for both Java and COBOL, which is not included in the previous figures.
In particular, code for invoking COBOL services from Java and vice-versa is fully generated.
Modeling is currently performed with a graphical, UML-based modeling tool.

The majority of the COBOL programs is still run on mainframe computers, both in traditional batch processing as well as interactive applications.
The remaining COBOL programs run on Linux servers.
Java EE applications are run on traditional Java application servers; some components are also deployed within an OSGi container on the mainframe.

The current release process mostly follows a traditional waterfall model with one major release every three months.
As a consequence, it usually takes several months to ship a new feature into production.
It should be noted that most of these applications are only used by professional users, who strongly prefer a stable environment to work in.
Therefore, a high rate of change is not necessarily desirable in this setting.

However, the current process is particularly cumbersome for changes that need to be implemented at a specific point in time, such as legal requirements.
Such changes are quite common in the financial services industry.
Obviously, such a change needs to be part of the last release \emph{before} the target date, and is therefore usually subject to a considerably higher delay.
This can be challenging with requirements that come at short notice.

Furthermore, it has been observed that the applications evolve at different speeds.
On the one hand, there are very mature applications that change seldomly, but irregularly, e.g., to keep up with changes in regulations.
Other applications, on the other hand, need to evolve frequently.
The fixed release schedule does not really match the needs of both these application types, and frequently causes friction.

In order to address these issues, it was decided to move towards a more flexible release model, and, in particular, to enable the individual applications to be deployed independently.
However, while the underlying technology platforms provide facilities for dynamic re-deployment, the applications themselves are currently too tightly coupled to allow for independent evolution.

There are two major reasons for this tight coupling.
The first is due to the fact that COBOL programs operate on \qt{bare} memory, and therefore require all data to conform to an exact memory layout to function properly.
Although this largely also applies to other languages such as C, COBOL lacks language features (such as user-defined functions) to properly abstract from this layout.
Furthermore, all data fields (including character strings and tables) have static sizes in COBOL.\footnote{There are some exceptions, such as unbounded tables. However, to our knowledge, these features are seldomly used in practice.}
As a consequence, even changing the length of a string field breaks the memory layout and thus requires to recompile and redeploy all affected programs.

The second reason results from the way code generation is currently handled.
As previously mentioned, client access code is generated automatically.
This code generation is part of the provider build, and the resulting artifacts are published as libraries for use by the clients.
While this approach is convenient for the developers, it leads to a significant sharing of code between applications.
Such sharing can (and, often enough, does) cause problems, especially when two embedded libraries rely on incompatible versions of a third library.

\section{A Simple Evolution Example}
\label{sec:example}

\noindent In order to better illustrate our context of evolution, consider a simple example.
Note that, for the sake of brevity, we omit common attributes like IDs.

\begin{figure*}[tb]
\begin{center}
\begin{subfigure}[b]{0.4\textwidth}
\begin{center}
\includegraphics[scale=0.45]{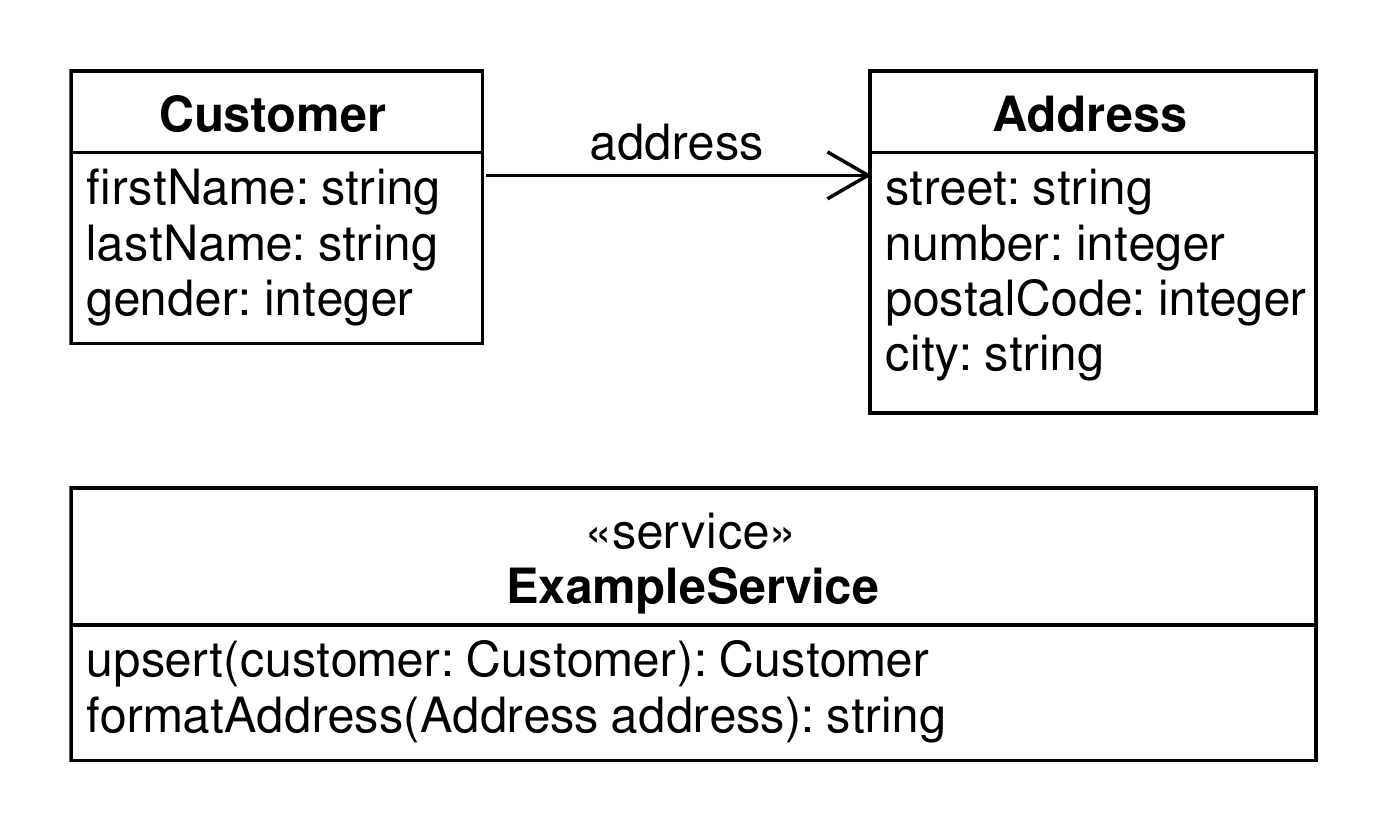}
\end{center}
\caption{Initial API model}
\label{fig:exaple-initial}
\end{subfigure}
\rulesep
\begin{subfigure}[b]{0.55\textwidth}
\begin{center}
\includegraphics[scale=0.45]{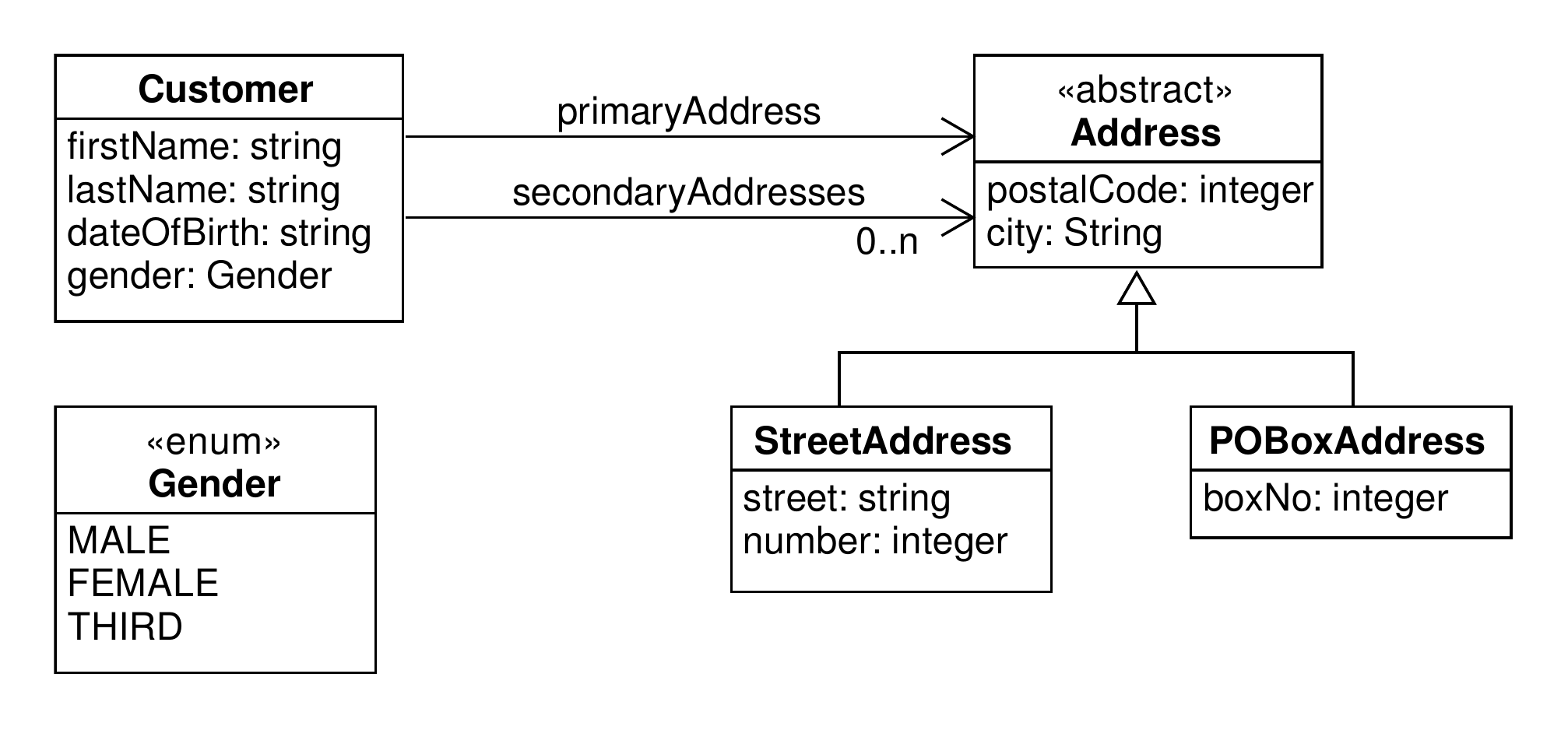}
\end{center}
\caption{Evolved API model}
\label{fig:exaple-final}
\end{subfigure}
\end{center}
\caption{UML model of the evolution example}
\end{figure*}

An application that manages customer data provides an API consisting of two types, as depicted in Figure~\ref{fig:exaple-initial}: 
The \code{Customer} type consists of the customer's first and last name, its gender, and its postal address.
The latter is modeled by the second type, \code{Address}, that represents a typical street address (street name, number, city name, and postal code).
As the API was inspired by an old (i.e., pre-Java 5) Java class structure, the customer's gender is modeled as an integer instead of an enumeration.
The API provides two service operations.
One inserts or updates a customer with given data (\qt{upsert}), and the other formats an address according to the rules prescribed by the local postage service.

Due to requests from client applications, the following changes are made to the API:

\begin{enumerate}
\item \emph{Field Addition:} The customer's date of birth, which is already present in the existing data, is added to the \code{Customer} type.

\item \emph{Field Renaming:} As customers may have secondary addresses, a new list field \code{secondaryAddresses} is added to the \code{Customer} type. As part of this change, the \code{address} field is renamed to \code{primaryAddress}.

\item \emph{Type Change:} A \code{Gender} enumeration is created, and the type of the field \code{gender} is changed accordingly.

\item \emph{New Enum Constant:} A third gender is added to the \code{Gender} enumeration.

\item \emph{New Specialization:} Support for P.O. box addresses is added. These consist of a box number, a postal code, and a city name. Therefore, the former \code{Address} type is renamed to \code{StreetAddress}, a new type \code{POBoxAddress} is created, and the common fields are moved to a common a supertype \code{Address}.

\end{enumerate}

\noindent The resulting API is shown in Figure~\ref{fig:exaple-final}.
In the following section, we will investigate how these changes could be realized using the state of the art, and highlight the shortcomings we intend to address.

\section{API Evolution}
\label{sec:existing-approaches}

\subsection{API Evolution Concepts and Patterns}
\noindent The main goal of API evolution is to apply changes to APIs without breaking the clients, i.e., the clients can remain unchanged and will still work as intended.
A fundamental pattern to allow for such changes is the \emph{tolerant reader pattern} \cite{Daigneau2012}.
The core idea of this pattern is that an application (\emph{reader}) that receives data from another application (\emph{writer}) does not require the data to exactly match its expectation, but tolerates certain deviations.
For example, additional data fields can usually be simply and safely ignored by the reader.

In order to identify such additional fields, however, the reader requires knowledge about the underlying schema of the data, such as field names.
Therefore, data formats with embedded field names are commonly used, e.g., XML or JSON.
And dynamic query facilities like XPath provide convenient options for creating tolerant readers.
However, the field names make these data formats quite verbose.
This has led to more optimized formats like Apache Avro, which is described below.

While tolerant readers allow for the safe addition of new fields by a writer, additional measures are required to support other changes, such as renaming a field.
In such cases, the tolerant reader can be complemented by a \emph{magnanimous writer}, i.e., a writer that provides more data than necessary.
For instance, a renamed field can be written both under its new and its old name, and since the readers ignore unexpected fields, neither old nor new readers break.

Deletions of fields or type changes are more difficult.
Mandatory fields cannot be deleted as long as there are still readers that expect them.
Type changes would require to write two different values \emph{under the same name}.
As a consequence, these changes cannot be addressed by tolerant readers and magnanimous writers alone, but require more elaborate evolution patterns.
A selection of common patterns is described in \cite{LuebkeEtAl2019}.
In the following paragraphs, we refer to these unless noted otherwise.

A common strategy to address potentially breaking changes like field deletions is to introduce \emph{version identifiers}, often based on \emph{semantic versioning}.\footnote{\url{https://semver.org}}
Semantic versioning expects that a version number is of the form $\textit{major}.\textit{minor}.\textit{patch}$, where the major version number is incremented if and only if an incompatible (i.e., breaking) change is made.

To make use of versioning, each client must provide the identifier of the desired API version when initiating an interaction with a provider.
Different approaches for accessing specific API versions are employed in practice, e.g., specific URLs or custom HTTP headers \cite{Sturgeon2016}.
It is then the provider's responsibility to proceed according to the given version, or to refuse the interaction if it is unable to do so.
Obviously, a provider must support at least two, sometimes even more (Pattern \emph{Two in Production} or \emph{$N$ in Production}, respectively), versions concurrently for a certain amount of time to avoid breaking clients.

Supporting multiple versions concurrently can be tedious.
Therefore, providers try to keep the number of such versions low.
A common pattern is to give a \emph{Limited Lifetime Guarantee}, which means that already upon publication, the provider announces until when this particular version will be supported. 
A slightly different approach is \emph{Aggressive Obsolescence}.
Here, the provider declares an API as deprecated at a time of his own choosing (e.g., when a new version becomes available), and gives the clients a sufficient grace time to migrate to a supported version.

Such processes cannot be reliably managed without tool support.
Registries keep track of published APIs and manage their life cycles \cite{Medjaoui2019}.
Furthermore, such registries may provide additional functionality like usage monitoring for operations as well as monetarization.

\subsection{Evolvability in Existing RPC Implementations}

\noindent After establishing the basic patterns for API evolution, we now proceed to investigate how evolution is supported in two common RPC implementations. We chose Apache Thrift because its evolution capabilities are well documented, and Avro due to its dynamic schema matching, which was a major inspiration for our approach.

\subsubsection{Apache Thrift}
Originally created at Facebook, Apache Thrift\footnote{\url{https://thrift.apache.org/}} is a cross-platform RPC framework that supports numerous target languages.
Thrift APIs are built from five major elements: 

\begin{itemize}
	\item \emph{Base types}, such as binary integers or character strings,
	\item \emph{Structures} composed of fields with types defined within the API,
	\item \emph{Containers}, such as lists or sets,
	\item \emph{Services}, which describe the operations provided by the API in terms of the defined types, and
	\item \emph{Exception types}, which are effectively structures representing an error
\end{itemize}

Thrift provides an own interface definition language (IDL) to specify APIs.
From such an API specification, a code generator produces the required source code artifacts for serializing and deserializing data.

Thrift's evolution capabilities are discussed in \cite{SleeAgarwalKwiatkowski2007}.
They are based on field identifiers and type specifiers:
At code generation time, each field receives a unique numeric id that is encoded along with its according type specifier and the actual value.
Although field identifiers can be assigned automatically, it is considered good practice to specify them explicitly.

The use of field identifiers makes Thrift definitions resilient to name changes and field reordering.
The type specifier ensures that the type of each encoded value is always known, even (and especially) for unexpected fields.
This, together with the fact that all types in Thrift are encoded in a self-delimiting way, allows to skip unexpected fields without error.

With respect to our evolution example, Thrift supports the field addition (Change 1) and the renaming (Change 2) very well.
This includes a non-obvious detail of the field addition due to the request-response communication.
As the field is added to a type that is used as an input parameter, the provider must not expect this field to be supplied, as older clients are not aware of it.
To account for this, Thrift fields are, by default, optional for input (opt-in), i.e., they are optional for the client, but mandatory for the provider.

The type change (Change 3) must be implemented by adding a new field with a different name.
Additional enum members can be freely added; readers treat unknown values as if no value had been provided at all.

Thrift provides no immediate support for inheritance, but it can be emulated by means of unions.
Unions in Thrift are special records with multiple fields, only one of which has a value.
As the entire construct needs to be manually emulated, there is no particular support for Change 5.

\subsubsection{Apache Avro}
Apache Avro\footnote{\url{http://avro.apache.org}} is a data serialization system that was originally developed in the context of Apache Hadoop, a platform for large-volume batch data processing.
Although primary a serialization system, Avro also specifies an RPC protocol.

Avro is particularly interesting for evolution as it employs dynamic schema resolution.
Unlike Thrift (and many other such frameworks), Avro does not rely on generating code from a common IDL definition; it only requires that both reader and writer have compatible schemas.
The Avro RPC protocol contains a handshake mechanism so that client and provider can agree on schemas before invoking a procedure.

Serializing and deserializing in Avro roughly works as follows.
The writer serializes the input using its own schema, and supplies the schema definition (or a reference to it) along with the serialized output.
The reader then resolves its own schema against the writer's schema, identifying potentially matching elements by name.
Schema evolution is enabled by allowing certain deviations between the two schemas \cite{Avro2020}.
For instance, a reader may specify alias names for writer fields, and certain types can be promoted (e.g., \code{int} can be promoted to \code{long}).

The result of the schema resolution is essentially a partial mapping from the writer's schema to the reader's schema.
This mapping is then applied to transform the data, which is deserialized using the writer's schema, to match the reader's expectations.

This schema resolution has three important consequences.
First, it removes the need to embed field identifiers in the actual payload data (\qt{untagged data}) and can therefore be applied to a wider variety of data formats.
Second, it allows for provider and client schemas to be specified independently, as long as they are sufficiently compatible.
Third, the explicit client schema enables the provider to deliver the data according to the needs of the client.
This is crucial for contexts where rigid and predefined data formats must be used, such as COBOL programs.
Therefore, our approach uses a very similar mechanism that only provides additional features to account for the static nature of COBOL.

It is apparent from the aforementioned examples that Avro's evolution capabilities primarily facilitate \emph{backwards compatibility}, i.e., enable a newer reader to read data written by an older writer.
To actually achieve backward or forward compatibility, certain rules must be actively followed by the developer, such as not adding or removing fields without default values \cite{Kleppmann2017}.

As for our evolution example, Avro supports the field addition (Change 1) in a way similar to Thrift.
However, the new field needs to be explicitly specified as optional, since Avro does not provide an equivalent concept to Thrift's opt-in.
The name change (Change 2) is partially supported by aliases.
But as these are only applied when reading, the (unchanged) client may not have an appropriate alias when reading the response.
The type change (Change 3) also needs a new field in Avro; but Change 4 requires the new enum to specify an explicit default value that is used in case of an unexpected value.
Similar to Thrift, there is no inheritance mechanism in Avro, but again, it can be emulated with union types.

Avro is extensively used in stream processing platforms such as Confluent or Apache Pulsar.\footnote{\url{http://pulsar.apache.org/}}
As such platforms distribute immutable data, they match Avro's reader-writer model very well.
In order to keep readers and writers compatible, both Confluent and Pulsar provide a schema registry, which manages the life cycle of the schemas and is capable of detecting (and possibly rejecting) changes that break the desired level of compatibility.\footnote{\url{https://docs.confluent.io/platform/current/schema-registry/avro.html}}\footnote{https://pulsar.apache.org/docs/en/schema-evolution-compatibility/}

\subsection{Summary}
\noindent In conclusion, several of the changes of our example are already quite well supported by existing implementations.
But especially more complex operations like type changes must be performed manually, which is potentially error-prone.
Furthermore, when using semantic versioning, it is the developer's responsibility to manually identify breaking changes and adjust the version number accordingly.
This is, however, far from trivial, especially in complex APIs.
In practice, while semantic versioning is more and more adopted, it has been found that breaking changes are commonly introduced also in minor and patch revisions \cite{RaemaekersVanDeursenVisser2014}.
\section{Our Approach to API Evolution}
\label{sec:approach}

\subsection{Overview}
\noindent The central idea of our approach is not to have schemas explicitly modeled by the developer (or API designer), but to derive these as well as the internal representations from a set of supported API definitions.
These definitions always exist as part of a revision history.
Thus, we always have access to adjacent revisions (provided that they exist), and are able to specify evolution steps from the previous revision directly in the API definition.
This way, we can -- at least to a large extent -- automatically ensure that the required rules are followed, and do not need to burden the developer with them.

\begin{figure}[bt]
\begin{lstlisting}[language=apidef]
record Customer {
 string firstName
 string lastName
 string dateOfBirth  
 Address primaryAddress replaces address
 Address* secondaryAddresses
 Gender gender as genderNew
}
\end{lstlisting}
\caption{Example provider API definition with evolution}
\label{fig:example-provider-definition}
\end{figure}

To illustrate the general approach, consider the example shown in Figure~\ref{fig:example-provider-definition}; the definition language is described in more detail later on.
In the example, we see an excerpt from an API definition representing the first three changes to the \code{Customer} type in our running evolution example.

First, recall that in its initial state, the \code{Customer} type had four fields, namely \code{firstName}, \code{lastName}, \code{address}, and \code{gender}.
The latter was modelled as an \code{integer}.
The first two fields still exist with the same name and the same type, and are therefore considered to be the same fields.
The \code{dateOfBirth} field is identified as a new field, since no matching field exists in the previous revision (Change 1).
Similarly, the \code{primaryAddress} and \code{secondaryAddress} fields do not have a match; however, the \code{replaces} clause reveals \code{primaryAddress} to be the same field as \code{address} from the previous revision (Change 2).
The \code{gender} field does exist in the previous revision, but with a different type (note that due to the same name, no explicit \code{replaces} clause is required).
Therefore, it is considered a type change, resulting in an implicit deletion of the previous field and the creation of a new one, but with the same name (Change 3).

The type change requires special treatment by the provider code.
As long as not all clients have updated to the respective revision, the provider must be able to handle the old field as well as the new field.
Therefore, the provider-internal representation of the type, e.g., the corresponding Java class, must contain both fields, although only one of them will be provided by or returned to a given client.
To avoid name clashes in the internal representation, the \code{as} clause allows us to provide an internal name of our choice.
This internal name is, of course, not part of the external API.
By always assigning new internal names to new fields, we can minimize the required changes to existing code.
In the given example, the old \code{gender} field will remain almost unchanged in the internal representation.
It will, however, become optional, as newer clients will no longer provide it.

On the client side, we also use an explicit API definition, albeit more simple.
Client API definitions are not revisioned; instead, they always refer to a specific provider API revision.
Therefore, they cannot contain \code{replaces} clauses.
Mapping public to internal names, however, is also possible in client definitions.
This way, we can prevent name changes from trickling into the client code.
An excerpt of an example client API definition is shown in Figure~\ref{fig:example-client-revision}.

\begin{figure}[bt]
\begin{lstlisting}[language=apidef]
record Customer as Person {
 string firstName
 string lastName as familyName
 integer gender
}
\end{lstlisting}
\caption{Example client API definition}
\label{fig:example-client-revision}
\end{figure}

The process of invoking a service can be summarized as follows.
First, the client encodes the request data according to a schema derived from its local API definition.
The definition itself, or at least a reference to it, is passed along with the data.
The provider then matches the client definition against the referenced revision; a process similar to Avro's schema resolution.
It then proceeds to decode the given data using the schema derived from the client's definition, and transforms it into the appropriate internal representation.
After returning from the service operation, this process is done in reverse order.
By using the client's definition to encode the result, we ensure that the response can be processed by the client.

Before looking into the details of the approach, we take a quick glance at Changes 4 and 5 of our running example.
We support both adding new enum members and sibling types, but these changes cannot be handled fully transparently.
Because both introduce new, valid values that cannot be represented in earlier API revisions, but at existing locations.
For instance, a new member in an existing enumeration (Change 4) can occur at any existing field of this type.
The addition of the new address type (Change 5) is similar:
Although the provider still supports the old \code{primaryAddress} field, it may be unable to determine a valid value, since a customer may now choose a \code{POBoxAddress} as its primary address.
Therefore, an explicit treatment of unrepresentable values is required.

\begin{figure}[bt]
\begin{center}
	\includegraphics[scale=0.5]{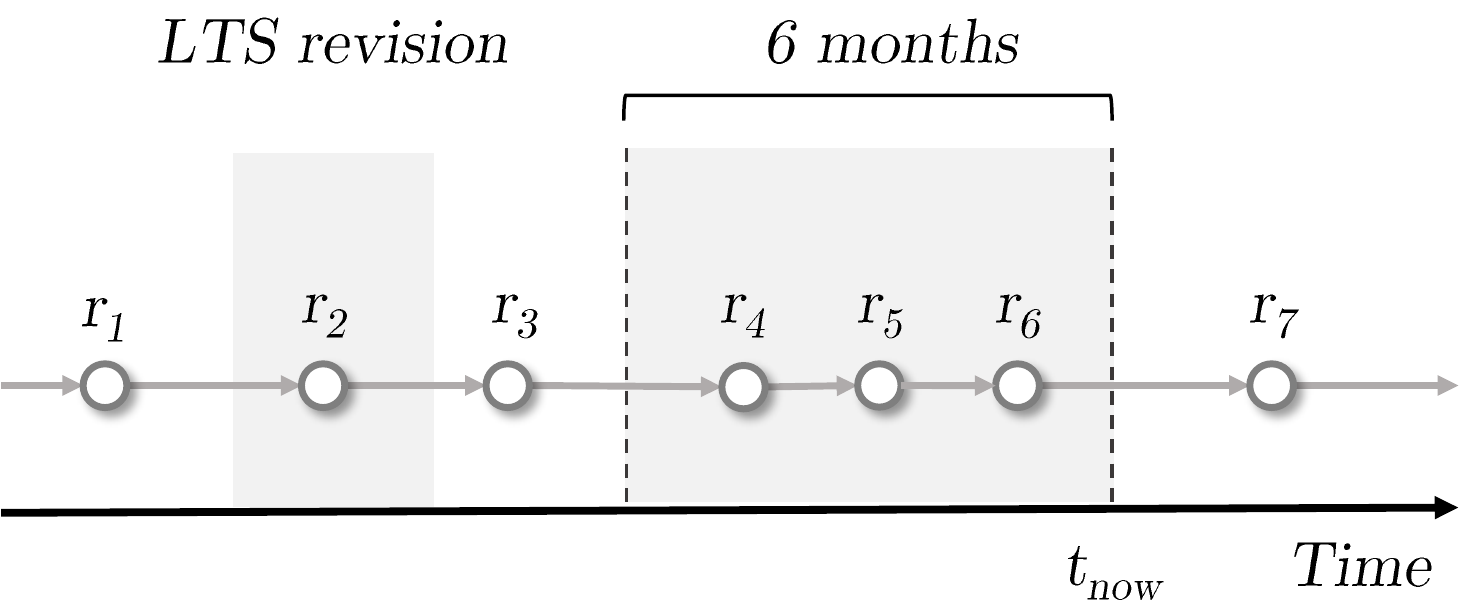}
\end{center}
\caption{Revision history with supported revisions}
\label{fig:revision-history}
\end{figure}

As discussed in the previous section, explicit API versioning is usually employed for breaking changes such as the removal of a field.
We take a different approach:
Each change (set) to an API definition leads to a new revision, whether breaking or not.
The revisions are stored in an API registry.
The provider then selects a set of supported revisions from the revision history, and our approach derives the necessary structures to support these revisions.
As a consequence, removing a field in a new revision does not mean that the field is removed immediately, but represents the \emph{intent} to remove this field.
While clients aware of the new revision will never receive it, the provider is still required to support the field until the last revision containing it is removed from the supported revision set.
To support this process, the registry also keeps track of client definitions.
The supported set approach allows to combine the previously presented lifecycle patterns in various ways.
Consider, for instance, the situation depicted in Figure~\ref{fig:revision-history}.
In this example, the provider supports all new revisions for six months (revisions $r_4$ to $r_6$), and additionally provides long-term support for revision $r_2$.

\begin{figure*}[bt]
\begin{center}
\includegraphics[scale=0.49]{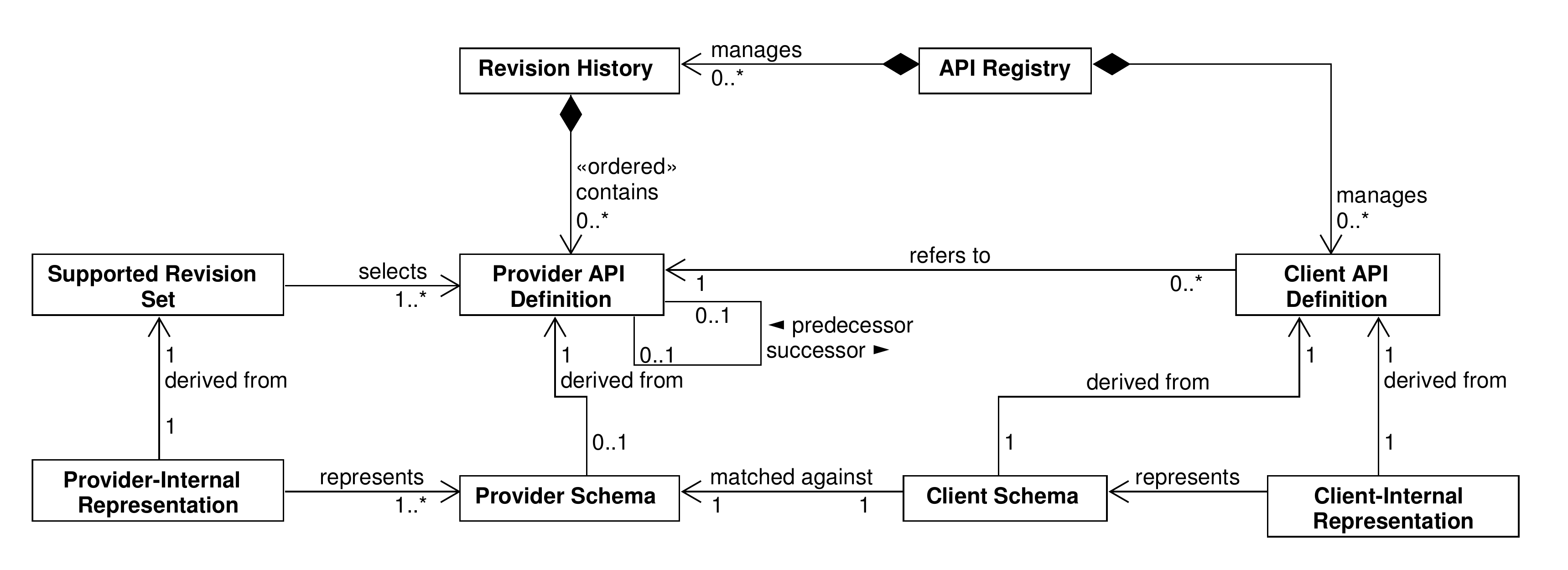}
\end{center}
\caption{Core elements of the approach}
\label{fig:core-elements}
\end{figure*}

The central elements of this approach and their connections are summarized in Figure~\ref{fig:core-elements}.
Selected elements of the approach are described in detail in the following paragraphs.

\subsection{The API Description Language}
\noindent The first aspect of our approach that we will describe in more detail is the specification of API definitions.
Similar to Thrift definitions and Avro protocols, an API definition consists of a number of service operations, together with a description of their input and output types.

\begin{figure*}
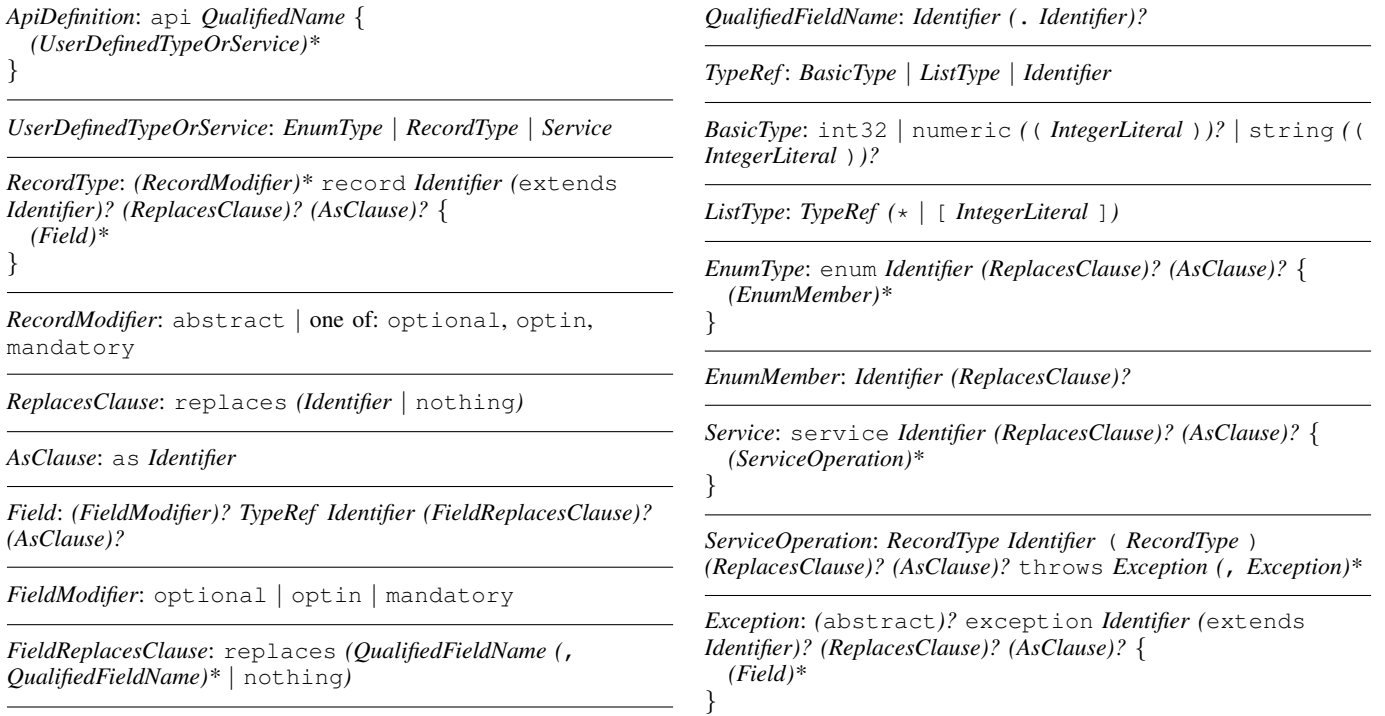

\begin{center}
\small
\begin{multicols}{2}
\begin{raggedright}

\nonterminal{ApiDefinition}: \terminal{api} \nonterminal{QualifiedName} \terminal{\{}\\
\quad \multiTerm{\nonterminal{UserDefinedTypeOrService}}\\
\terminal{\}}

\ruleseparator

\nonterminal{UserDefinedTypeOrService}: \nonterminal{EnumType} $|$ \nonterminal{RecordType} $|$ \nonterminal{Service}\\

\ruleseparator

\nonterminal{RecordType}: \multiTerm{\nonterminal{RecordModifier}} \terminal{record} \nonterminal{Identifier} \optionalTerm{\terminal{extends} \nonterminal{Identifier}} \optionalTerm{\nonterminal{ReplacesClause}} \optionalTerm{\nonterminal{AsClause}} \terminal{\{}\\
\quad \multiTerm{\nonterminal{Field}}\\
\terminal{\}}\\

\ruleseparator

\nonterminal{RecordModifier}: \terminal{abstract} $|$ one of: \terminal{optional}, \terminal{optin}, \terminal{mandatory}\\

\ruleseparator
\nonterminal{ReplacesClause}: \terminal{replaces} \parenTerm{\nonterminal{Identifier} $|$ \terminal{nothing}}\\

\ruleseparator
\nonterminal{AsClause}: \terminal{as} \nonterminal{Identifier}\\

\ruleseparator

\nonterminal{Field}: \optionalTerm{\nonterminal{FieldModifier}} \nonterminal{TypeRef} \nonterminal{Identifier} \optionalTerm{\nonterminal{FieldReplacesClause}} \optionalTerm{\nonterminal{AsClause}}

\ruleseparator

\nonterminal{FieldModifier}: \terminal{optional} $|$ \terminal{optin} $|$ \terminal{mandatory}

\ruleseparator

\nonterminal{FieldReplacesClause}: \terminal{replaces} \parenTerm{\nonterminal{QualifiedFieldName} \multiTerm{\terminal{,} \nonterminal{QualifiedFieldName}} $|$ \terminal{nothing}}

\ruleseparator

\nonterminal{QualifiedFieldName}: \nonterminal{Identifier} \optionalTerm{\terminal{.} \nonterminal{Identifier}}

\ruleseparator

\nonterminal{TypeRef}: \nonterminal{BasicType} $|$ \nonterminal{ListType} $|$ \nonterminal{Identifier}

\ruleseparator

\nonterminal{BasicType}: \terminal{int32} $|$ \terminal{numeric} \optionalTerm{\terminal{(} \nonterminal{IntegerLiteral} \terminal{)}} $|$ \terminal{string} \optionalTerm{\terminal{(} \nonterminal{IntegerLiteral} \terminal{)}}\\

\ruleseparator

\nonterminal{ListType}: \nonterminal{TypeRef} \parenTerm{\terminal{*} $|$ \terminal{[} \nonterminal{IntegerLiteral} \terminal{]}}

\ruleseparator

\nonterminal{EnumType}: \terminal{enum} \nonterminal{Identifier} \optionalTerm{\nonterminal{ReplacesClause}} \optionalTerm{\nonterminal{AsClause}} \code{\{}\\
\quad \multiTerm{\nonterminal{EnumMember}}\\
\terminal{\}}\\

\ruleseparator

\nonterminal{EnumMember}: \nonterminal{Identifier} \optionalTerm{\nonterminal{ReplacesClause}}\\

\ruleseparator

\nonterminal{Service}: \terminal{service} \nonterminal{Identifier} \optionalTerm{\nonterminal{ReplacesClause}} \optionalTerm{\nonterminal{AsClause}} \terminal{\{}\\
\quad \multiTerm{\nonterminal{ServiceOperation}}\\
\terminal{\}}\\

\ruleseparator

\nonterminal{ServiceOperation}: \nonterminal{RecordType} \nonterminal{Identifier} \terminal{(} \nonterminal{RecordType} \terminal{)} \optionalTerm{\nonterminal{ReplacesClause}} \optionalTerm{\nonterminal{AsClause}} \terminal{throws} \nonterminal{Exception} \multiTerm{\terminal{,} \nonterminal{Exception}}\\

\ruleseparator

\nonterminal{Exception}: \optionalTerm{\terminal{abstract}} \terminal{exception} \nonterminal{Identifier} \optionalTerm{\terminal{extends} \nonterminal{Identifier}} \optionalTerm{\nonterminal{ReplacesClause}} \optionalTerm{\nonterminal{AsClause}} \terminal{\{}\\
\quad \multiTerm{\nonterminal{Field}}\\
\terminal{\}}\\
\end{raggedright}
\end{multicols}
\normalsize
\end{center}
\caption{Grammar overview of the API definition language}
\label{fig:adl-grammar-overview}
\end{figure*}

An overview of the grammar of our API definition language, which we already used in the previous examples, is given in Figure~\ref{fig:adl-grammar-overview}.
Note that for the sake of brevity, we do not give definitions for some non-terminals that are obvious or will become obvious by the examples below.
An \emph{API definition} consists of a number of user-defined types (i.e., record types and enumeration types) and services.

\emph{Enumeration types} consist only of named values; unlike, for instance, Java enumerations, we do not support fields on members of an enumeration.
\emph{Record types} comprise named fields of arbitrary types, as described below.
In order to facilitate modeling records with common fields, such as the addresses in the running example, a simple inheritance mechanism exists on record types.
However, no overloading is supported.
Record types can be declared \code{abstract} or given an optionality modifier, which serves as the default for the fields declared within the type.
The optionality modifier is inherited to the record's subtypes, but can be overridden by them.

\emph{Fields} of a record can be of a user-defined type, a basic type (currently 32-bit binary integers, numeric strings and regular character strings), or an ordered list type.
To account for languages like COBOL, all types are either bounded by construction (e.g., \code{int32}, records, or enums), or can be given explicit bounds, such as \code{string(5)} for a string of at most length 5 or \code{string(5)[10]} for a list of at most 10 strings of at most length 5.\footnote{Note that the different types of brackets are required for syntactic unambiguity.}
An optionality modifier can be defined for each field, which can be either \code{optional} (the field is optional for both in- and output), \code{optin} (the field is optional for input, but mandatory for output) or \code{mandatory}.
If no optionality is specified, the containing type's default optionality is used.

\emph{Services} essentially serve as named containers for service operations.
Each \emph{service operation} has one record type as its input type, and one record type as its output type.
It may furthermore throw exceptions, where an \emph{exception type} is a record type that is restricted to be used only at this location.

\paragraph*{Public and Internal Names}
Every user-defined type, field, service, and service operation must have a \emph{public name}, which is used for schema generation and matching.
An addition, each such element can be assigned an \emph{internal name} by means of an \code{as} clause, which is only used for the internal representation.
While the public name must only be unique within the respective scope of the current API definition, the internal name must be unique in all supported revisions.
If no explicit internal name is specified, the public name is used.

\paragraph*{Replacement Clauses}
All of the previously described elements can be used in both client and provider API definitions.
An element exclusive to provider API definitions is the \code{replaces} clause.
Such a clause may be added to any element with a public name, and denotes the element \emph{in the previous revision} that should be considered as the predecessor of the given element.
In most cases, this element is identified only by its name; in the case of fields, this name may be further qualified by a type name.
If no explicit replacement is given, but a matching element of the same name exists in the previous revision, it is implicitly assumed to be the predecessor.
This can be suppressed by the specifying that the given element \code{replaces nothing}.
The semantics of replacement are discussed in detail in the following section.

\subsection{API Evolution}
\noindent The fundamental means for evolving APIs in our approach is to establish relations between the elements of consecutive API revisions using the replacement clauses of our API definition language.
In total, this results in five relations, namely one on types, fields, enum members, services, and service operations.
These relations define which elements are considered to be the same across revisions, and play a key role in transforming representations.
Unrelated elements from the \qt{old} revision are considered deleted, and unrelated elements from the \qt{new} revision are considered added.
In order for the transformations to work, we require these relations to be injective partial functions.
The reason for this requirement is given in the section on representation transformation below.
Therefore, we must ensure that no two elements share the same predecessor.

Unless explicitly suppressed, we assume that two elements from consecutive revisions are related if they have the same name and are compatible (implicit replacement).
To be compatible, the elements must be of the same kind (e.g., record type, enum type, or service).
For fields, we furthermore require the field's type to be related, and for service operations, both the input and the output type must be related.
Anonymous types, such as lists, are considered related if their referenced types are related.
Furthermore, we require bounds on bounded types to match.
If an explicit replacement is specified, the element with the given name is assumed instead, and the same rules for compatiblity apply.

To allow for type changes, replacements for fields and service operations may also refer to elements that are incompatible with respect to types.
These elements, however, are not related.
Therefore, this results in the removal of the old and the addition of the new element, possibly under the same name.
To further illustrate these rules, several examples are given in Table~\ref{tab:example-evolution-steps}.

\begin{table*}[bt]
\begin{center}
\begin{tabular}{|l|l|l|}
\hline
\textbf{Previous revision} & \textbf{Current Revision} & \textbf{Description}\\
\hline
\hline
\code{\textbf{record} A} \{ & \code{\textbf{record} B \textbf{replaces} A} \{ & Record type \code{A} is renamed to \code{B}\\
$\;$ \code{string a} & $\;$ \code{string d \textbf{replaces} a} & Field \code{a} in \code{A} is renamed to \code{d}\\
$\;$ \code{int32 b} & $\;$ \code{numeric(5) b} & Type change for \code{b} from \code{int32} to \code{numeric(5)}\\
\code{\}} & $\;$ \code{int32 c \textbf{replaces} b} & Error, multiple replacements for \code{b}\\
& $\;$ \code{string y \textbf{replaces} x} & Error, \code{x} does not exist in \code{A}\\
& $\;$ \code{string z} & New field \code{z}\\
& \code{\}} & \\
\code{\textbf{record} X \{} & & Type \code{X} is deleted\\
$\;$ \code{string x} & & \\
\code{\}} & & \\
\hline
\end{tabular}
\end{center}
\caption{Examples for valid and invalid evolution steps}
\label{tab:example-evolution-steps}
\end{table*}

The inheritance mechanism of record and exception types provides additional options for evolution.
Fields may be pulled up or pushed down along the inheritance hierarchy.
However, we deliberately do not support changing an existing supertype of a type.
While simple on the schema level, this type of change makes the internal representation and thus the programming model significantly more complex.

Before we discuss these changes in detail, we need to take a quick look at the semantics of inheritance in our approach.
As previously noted, our inheritance mechanism only provides basic subtyping, and facilitates the modeling of structures with common fields.
Internally, each subtype receives an individual copy of each inherited field, and the supertype is replaced by a union type consisting of all possible subtypes.
If the supertype is not abstract, it is added to the union as well.
This interpretation greatly facilitates handling inheritance-related evolution, as shown below.

Recall from the running example that in the last change, a new sibling type to the street address was created, and that the common attributes were moved to a new supertype.
This change is best modeled by declaring \code{StreetAddress} as the successor to \code{Address}, and then pulling up the fields to the new supertype \code{PostalAddress}.
We deliberately allow to pull up fields that are not present in all subtypes (otherwise, we would first have to publish a revision with \code{POBoxAddress} having them), but this requires us to explicitly specify the type from which we are pulling.
We also allow to pull up multiple fields into one, as long as they have the same type.

As opposed to pulling up, a field to be pushed down is uniquely determined by its name.
But in order to clearly separate a push-down from a simple replacement, we have chosen to require the field to be qualified with the respective supertype's name.
A field can be pushed down to multiple subtypes, for instance, if it should remain in two of three possible types.
This is the only case in which it is legal to specify two successors to the same element.

Due to our interpretation of inheritance, these changes are not in conflict with our requirement of the predecessor relations being injective functions.
Especially pushing down a field to multiple subtypes appears to be a problem, but as each of the types has an individual instance of this field, these types are, in fact, unchanged.
The field is only removed from fields that no longer inherit it.
Conversely, pulling a field up only causes some types to receive a new field.
Several examples of such evolution steps are shown in Table~\ref{tab:example-evolution-steps-inheritance}.

\begin{table*}[bt]
\begin{center}
\begin{tabular}{|l|l|l|}
\hline
\textbf{Previous revision} & \textbf{Current Revision} & \textbf{Description}\\
\hline
\hline
\code{\textbf{abstract record} A} \{ & \code{\textbf{abstract record} A} \{ & \\
$\;$ \code{string a} & $\;$ \code{string a2 \textbf{replaces} B.b, C.c} & Pull-up of \code{b} and \code{c} into \code{A}\\
\code{\}} & $\;$ \code{string a3 \textbf{replaces} B.b2, C.c2} & Error, types do not match\\
\code{\textbf{record} B \textbf{extends} A} \{ & $\;$ \code{\}} & \\
$\;$ \code{string b} & \code{\textbf{record} B \textbf{extends} A} & \\
$\;$ \code{string b2} & $\;$ \code{string b3 \textbf{replaces} A.a} & Push-down of \code{a} into \code{B}\\
\code{\}} & \code{\}} & \\
\code{\textbf{record} C \textbf{extends} A} \{ & \code{\textbf{record} C \textbf{extends} A} \{ & \\
$\;$ \code{string c} & $\;$ \code{string c3 \textbf{replaces} A.a} & Push-down of \code{a} into \code{C}\\
$\;$ \code{int32 c2} & $\;$ \code{String c} & Error, multiple successors for \code{c}\\
\code{\}} & \code{\}} & \\
\hline
\end{tabular}
\end{center}
\caption{Examples for valid and invalid evolution steps with inheritance}
\label{tab:example-evolution-steps-inheritance}
\end{table*}

\subsection{Schemas and Internal Representations}
\noindent The API definition serves as the foundation of the schemas and internal representations for both clients and providers.

Schema derivation works exactly the same for clients and providers.
The process is very straight-forward, and is completely oblivious of the revision history.
We use a schema that is very similar to Avro's, the only (but -- at least for COBOL -- key) difference is the support for bounded lists and base types.
Conversely, if no bounded types are required, our approach can be easily adapted to Avro schemas.

Deriving the internal representation is considerably more complex, especially for the provider.
This is due to the fact that the provider's internal representation must be able to represent the entire supported revision set at the same time.
The client internal representation is essentially a provider representation derived from a singleton revision set.

In order to be able to represent the entire supported revision set, the internal representation must essentially be a union of all revisions in the set.
Depending on the capablities of the target language and the desired programming model, the precise algorithm differs slightly.

The rough outline is as follows.
First, all types and services from the latest supported revision are added using their internal names, including their respective sub-elements.
For each previous supported revision, all elements are added to the internal representation that do not have a successor in the next supported revision, i.e., have been deleted afterwards.\footnote{Recall that the supported revision set does not need to be contiguous.}
Elements that do have a successor are merged, in particular:

\begin{itemize}
\item If the newer revision of a type does not contain a field present in the earlier revision, the field is added

\item If the optionality of a field differs between revisions, the more permissive one is used

\item If the exceptions of a service method differ, the union of both methods is used
\end{itemize}

For target languages that support it, it may furthermore be desirable to actually use inheritance for the internal representation.
In such cases, a type must additionally be concrete if at least one of its supported revisions is concrete.

Obviously, the resulting internal representation contains every type, field, service, and service operation present in any of the supported revisions.
Note that all elements are based on their latest revision, e.g., will have their latest internal name.
Furthermore, the optionality of fields is chosen permissive enough for all revisions.

\subsection{Schema Matching and Representation Transformation}
As described in the overview, we employ schema matching similar to Avro's to decouple clients and providers as much as possible.
All matching and converting is done by the provider, so that the client can remain unchanged.

Schema elements are matched by name and must be of the same type; type promotion is not supported.
Since each client API definition is only matched against the specific provider revision it refers to, no precautions against renaming are necessary.
After decoding the request data using the client schema, the resulting mapping from the schema matching is used to associate the decoded values with the appropriate types and fields of the provider schema.
These, in turn, allow to identify the corresponding elements in the internal representation (e.g., the field to store the value in).
As the internal representation is always based on the latest revision of an element, but the match may involve an older revision (e.g., referring to an outdated internal name), this may require to navigate the successors of the respective element until the latest revision is found.

To encode the result from the internal representation, the process is performed the other way.
In particular, the predecessor relation may be required to associate the internal revision with the appropriate elements of the schema.
It is this bidirectional mapping that requires the predecessor relation to be an injective partial function.
Encoding the data is again performed using the client schema to ensure that the data is returned exactly as the client expects them.

\section{Evaluation}
\label{sec:implementation}

\noindent In order to assess the general feasibility of our approach, we conducted a short evaluation with respect to the goals stated in the introduction.
To address Goals~1 and 2, we conducted a small survey with nine professional developers from our case study.
We particularly selected participants that had several years of professional experience.
The participants were given a presentation of the approach in small groups, and were asked to fill out a short questionnaire afterwards.
Four of the respondents reported to use only Java in their everyday work, four both Java and COBOL, and one Java and JavaScript.

Regarding Goal 1, we asked the respondents to rate their agreement to three statements on a seven-point scale, ranging from \qt{fully disagree} (encoded as -3) to \qt{fully agree} (+3).

All participants agreed to the statement that the approach is comprehensible (fully agree: 4 / 5 / 0 / neutral: 0 / 0 / 0 / fully disagree: 0; median: 2).
The statement that the approach addresses a relevant problem was also agreed to, but was apparently seen quite differently among the participants.
While four participants fully agreed to this statement, three participants chose the neutral option (4 / 1 / 1 / 3 / 0 / 0 / 0; median: 2).
The participants were furthermore asked whether they could envision applying the approach in their everyday work, and whether it supported the types of changes they needed.
These statements were also generally agreed to (1 / 4 / 2 / 2 / 0 / 0 / 0; median: 2) and (2 / 1 / 3 / 2 / 1 / 0 / 0; median: 1), respectively.

Regarding Goal~2, we were particularly interested in identifying aspects of the approach that might negatively affect a developer's day-to-day work.
Overall, handling the potentially large number of optional fields in the provider code raised the greatest concern among the participants, with 6 participants agreeing to some degree to the statement that this might impede their everyday work (0 / 1 / 4 / 3 / 0 / 1 / 0; median: 1).
Creating and maintaining client schemas also led to some concern, but the respondents were mainly indifferent (0 / 0 / 1 / 4 / 1 / 2 / 0; median: 0).
Maintaining the provider histories (0 / 1 / 0 / 3 / 2 / 2 / 1; median: -1) and explicitly publishing them to a repository (0 / 0 / 1 / 1 / 2 / 3 / 2; median: -2) were not considered as notable impediments.

The respondents were furthermore able to add remarks in an open text field.
The most frequent remark was that a facility for highlighting semantic changes might be useful to make clients more aware of such changes.

We see the greatest threats to the validity of this survey in the size and structure of the sample, the sampling method, and potential biases.
Due to the small sample size and the fact that all participants work for the same company, the results may not be generalizable.
Several biases may result from the fact that the participants were selected by the researchers, and that one of the researchers, besides his research, works for the same company.
Therefore, the responses may be unduly favorable of the approach.

Despite these threats, we conclude that there is a strong indication that the approach is considered applicable in practice, although further research is necessary to support this indication.
As for Goal~2, the survey revealed no show-stoppers, but suggests that future work should also focus on programming patterns and tooling to efficiently handle the additional complexity for the provider.

To address Goals~3 and 4, we created a proof-of-concept implementation of our approach.
This implementation consisted of three components: One containing the API definition language and the schema resolution mechanism, and a runtime component for both Java and COBOL for performing the actual data conversions.
Since published API definitions are immutable, schema resolution is a one-time operation.
Therefore, the impact on runtime performance is determined by the conversion components.

In our experiments, we focused on two scenarios most relevant to our case study, namely remote invocations between Java applications and in-process calls between COBOL programs.
For Java, we used reflection-based decoders and encoders to convert the serialized client data directly into the provider's internal representation and vice versa.
A simple binary format was used for serialization.
In our experiments, the performance of our converters was comparable to that of the JSON libraries Jackson and Yasson.
For instance, converting a \code{Customer} object from client revision 1 to provider revision 6 and back took, on average, 1.67µs using our converters, while serializing and deserializing the object took 1.1µs with Jackson and 4.4µs with Yasson.

For our COBOL experiments, we created a converter program in C to serve as a proxy between two interacting COBOL programs and convert parameters and results as needed.
The transformations themselves were defined using a simple bytecode, which can by generated dynamically, and interpreted by the C program.
Although these conversions were considerably faster than the ones from our Java experiment, it must be noted that this is truly additional overhead, as COBOL parameters are usually just passed by reference.

Our proof-of-concept implementation suggests that our approach is indeed applicable to aged technologies like COBOL.
However, it may have a considerable performance impact due to additional overhead.
For remote communication, this is of less concern, since the conversions can be efficiently integrated into the necessary serializations and deserializations.
\section{Related Work}
\label{sec:related-work}

\noindent Evolving software as well as their data and artifacts in a compatible way is a major challenge, and a large body of research exists on various aspects of this topic.
The reasons for and challenges of API evolution have been investigated in multiple studies.
For instance, Dig and Johnson \cite{DigJohnson2006} report that the majority of breaking changes are caused inadvertedly by refactorings, and conclude that tool support is required to ensure compatibility with clients.
Research focused particularly on the evolution of public Web APIs reaches different conclusions.
While Fokaefs et~al. \cite{FokaefsEtAl2011} find that mostly compatible changes such as field additions are made, other studies report that client application developers suffer considerably from non-standardized or even no procedures for API evolution \cite{EspinhaZaidmanGross2014} and poor documentation and communication of changes \cite{SohanAnslowMaurer2015}.
Wang et~al. \cite{WangEtAl2014} observe that, in particular, the addition or removal of methods leads to many questions by client developers on Q\&A sites like StackOverflow.

One of the central aspects of our approach, namely deriving schemas and representations from modifications of API definitions, is closely related to \emph{co-evolution} or \emph{coupled evolution}.
A co-evolution approach that also uses traceability information from models to derive transformations on derived data is describe in \cite{VermolenVisser2008}.
The authors present an approach for specifying changes to an entity model, and use the traceability information to derive database migrations reflecting these changes.
Wagelaar et al. \cite{WagelaarEtAl2012} propose a virtual machine for executing transformation bytecode, which is a similar notion to the generic transformer for COBOL.

Compatible evolution has also been studied for binary artifacts.
The primary goal of binary compatibility is to allow that a newer version of, say, a library can be dynamically linked with a program that was compiled against an older version of this library.
For instance, a compiler may statically embed indexes of the virtual method table within the object code of a derived class.
If a new virtual method is added to the superclass, these stored indexes may become invalid, therefore breaking binary compatibility \cite{FormanEtAl1995}.

Most modern programming languages provide information about the changes that do and do not preserve binary compatibility.
The Java Language Specification \cite{GoslingEtAl2020}, for instance, contains an entire chapter on this topic.

\section{Conclusions and Future Work}
\label{sec:conclusions}

\noindent In this paper, we have presented an approach for continuous API evolution for enterprise software systems in order to support the independent deployability of single applications.
As shown by the proof-of-concept implementation, this approach is indeed viable and satisfies the goals described at the beginning of this paper.
Although it would certainly be possible to add more powerful evolution mechanisms, we decided to aim for an approach that is both easy-to-use and easy-to-understand for a developer.

In our future work, we plan to extend this approach in several ways.
First, we intend to address the findings and limitations of our evaluation.
Besides more sophisticated implementations, this includes tooling for managing, querying, and editing revision histories, and a more in-depth exploration of the resulting programming model.
Further opportunities for research include a more formal treatment of the API definition language and their underlying evolution semantics as well as the introduction of such concepts into existing API specifications like OpenAPI.

\end{document}